%% file: QCHS.tex
\documentclass{PoS}

\usepackage{amsmath}
\usepackage{bbold}
\usepackage[T1]{fontenc}

\title{Influence of quark masses
and strangeness degrees of freedom on inhomogeneous chiral phases}

\ShortTitle{Influence of quark masses and strangeness d.o.f.\ on inhomogeneous chiral phases}
\author{\speaker{Michael Buballa}%\thanks{A footnote may follow.}
       \\
        Theoriezentrum, Institut f\"ur Kernphysik, TU Darmstadt, Darmstadt, Germany\\
        E-mail: \email{michael.buballa@physik.tu-darmstadt.de}}

\author{Stefano Carignano\\
Departament de F\'isica Qu\`antica i Astrof\'isica and Institut de Ci\`encies del Cosmos, \\Universitat de Barcelona, Barcelona, Catalonia, Spain\\
        E-mail: \email{stefano.carignano@fqa.ub.edu}}

\abstract{
Several model calculations of the
 QCD phase structure at nonzero temperature and density suggest that in certain regions of the phase diagram inhomogeneous condensates are favored over homogeneous ones. 
 In particular, in a two-flavor NJL model in the chiral limit
 the would-be first-order boundary associated with the chiral phase transition 
 is entirely covered by an inhomogeneous phase, characterized by a spatially modulated chiral condensate, whose tip coincides exactly with the chiral critical point. 
In this contribution we discuss how this result is altered by the inclusion of nonzero quark masses and strange quarks in the model. 
}

\FullConference{XIII Quark Confinement and the Hadron Spectrum - Confinement2018\\
		31 July - 6 August 2018\\
		Maynooth University, Ireland}

\newcommand{\Eq}[1]{{Eq.~({\ref{#1}})}}

\newcommand{\ave}[1]{{\langle{#1}\rangle}}
\newcommand{\bea}{\begin{eqnarray}}
\newcommand{\eea}{\end{eqnarray}}
\newcommand{\beq}{\begin{equation}}
\newcommand{\eeq}{\end{equation}}
\newcommand{\beas}{\begin{eqnarray*}}
\newcommand{\eeas}{\end{eqnarray*}}

\def\p{{\bf p}}

\def\x{{\bf x}}

\begin{document}

\section{Introduction}

The possible existence of a chiral critical point in the QCD phase diagram at nonzero temperature $T$ and 
quark chemical potential $\mu$ has stimulated a large amount of theoretical and experimental efforts during the last two decades (see e.g. 
\cite{Stephanov:1999zu,Fodor:2004nz, Stephanov:2007fk, Friman:2011zz,Kumar:2013cqa}).
At low $T$ and nonvanishing $\mu$, where lattice QCD techniques cannot be applied,
QCD-inspired models, like the Nambu--Jona-Lasinio (NJL) model or the quark-meson (QM) model, typically find a first-order
chiral phase transition. Beyond a given temperature, this transition line eventually terminates at a critical endpoint (CEP) \cite{Asakawa:1989bq,Scavenius:2000qd},
above which only a crossover is found, in agreement with the lattice result for vanishing densities \cite{Aoki:2006we}.
 If these model calculations are performed considering only up and down quarks with vanishing bare masses (the so-called chiral limit)
one finds instead a second-order transition in the low-$\mu$-high-$T$ regime which is connected to the 
first-order phase boundary at low $T$ in a tricritical point (TCP).

The underlying assumption which has been used to obtain the aforementioned picture is 
 that the chiral order parameter is spatially homogeneous. 
If one instead allows for spatial modulations, inhomogeneous phases are found to be energetically 
favored over the homogeneous ones in certain regions of the phase diagram (for a review see \cite{Buballa:2014tba}).
 In particular, for a  two-flavor NJL model in the chiral limit one finds that the first-order phase boundary separating the homogeneous chirally broken
phase from the restored one is completely covered by an inhomogeneous phase. 
As a consequence, a Lifshitz point (LP), which is 
the point where three different phases (the homogeneous and inhomogeneous chirally broken phases,
together with the restored one) meet, appears in the phase diagram. 
In the simplest realization of a two-flavor NJL model in the chiral limit, it was shown 
 within a Ginzburg-Landau (GL) analysis that the LP
is located at the same position in the $T$-$\mu$ plane as the TCP in the case when inhomogeneous phases 
are not considered \cite{Nickel:2009ke}. 

In this contribution we discuss how the presence of bare quark masses and of strange quarks affects this result. 
Inhomogeneous phases in a two-flavor NJL model with nonzero current masses have already been studied in Ref.~\cite{Nickel:2009wj}.
There it was found numerically that the inhomogeneous region shrinks with increasing quark masses but still covers the entire
first-order phase boundary between the homogeneous chirally broken and restored phases. 
In Ref.~\cite{Buballa:2018hux}  we have recently confirmed this more rigorously within a GL analysis, 
showing that the tip of the inhomogeneous phase, which we termed ``pseudo-Lifshitz point'' (PLP), exactly coincides with the CEP.  
This will be discussed in section \ref{sec:masses}. 

In  section \ref{sec:strange} we extend our GL study to a three-flavor NJL model.
A first investigation of inhomogeneous phases in three-flavor matter has been performed in Ref.~\cite{Moreira:2013ura}, 
albeit only for vanishing temperatures and 
employing a simple explicit ansatz for the spatial dependence of the chiral condensate. Our analysis will instead allow us to investigate 
the relation between the CEP and the (P)LP, an aspect which is of particular interest: Indeed,
our motivation for including strange quarks is not only that they may play a role under realistic conditions, e.g., in astrophysical scenarios,
but also the fact that in the limit of three very light quark flavors the first-order phase boundary (and hence the
CEP) is expected to eventually reach the $T$ axis \cite{Stephanov:2007fk}. 
Therefore, if the (P)LP still coincides with the CEP for three flavors and assuming that the same holds in QCD, this particular limit would open 
the possibility to study inhomogeneous phases at $\mu=0$ on the lattice.

\section{Inhomogeneous phases away from the chiral limit}
\label{sec:masses}

We consider the standard NJL-model Lagrangian 
\beq
\mathcal{L} = \bar\psi\left( i\gamma^\mu \partial_\mu - m \right) \psi 
+ G \left\{ (\bar\psi\psi)^2 + (\bar\psi i\gamma_5 \vec\tau \psi)^2 \right\}\,,
\eeq
describing quark fields $\psi$ with 
bare mass $m$ and $N_f=2$ and $N_c=3$ color degrees of freedom,
interacting via scalar-isoscalar and pseudoscalar-isovector four-point vertices proportional to a coupling constant $G$.

We perform a mean-field approximation, linearizing the interaction in the presence of a scalar condensate 
$\ave{\bar\psi \psi}$. In order to describe inhomogeneous phases we generally allow the condensate to depend on the 
spatial coordinate $\mathbf{x}$ but we assume it to be time independent. 
For simplicity we restrict ourselves to scalar condensates. The inclusion of pseudoscalar condensates is straightforward 
but it is well known for homogeneous phases that they are disfavored against scalar condensates if $m\neq 0$.
In Ref.~\cite{Buballa:2018hux} we have shown that this is also true for inhomogeneous condensates, at least in the regime where the 
GL analysis we discuss below is valid.
The mean-field Lagrangian then takes the form
\beq
\mathcal{L}_\mathrm{MF} = \bar\psi\left( i\gamma^\mu \partial_\mu - M(\mathbf{x})\right) \psi 
- \frac{(M(\mathbf{x})-m)^2}{4G}\,,
\eeq
where
\begin{equation}
       M(\mathbf{x}) = m - 2G \ave{\bar\psi \psi}(\mathbf{x})
\end{equation}
can be interpreted as a space-dependent constituent quark mass.

Noting that $\mathcal{L}_\mathrm{MF}$ is bilinear in the quark fields, the mean-field thermodynamic potential per volume 
$\Omega(T,\mu)  = -T/V\, \log \mathcal{Z}(T,\mu)$
can be obtained by standard path-integral techniques. One finds
\beq
\Omega(T,\mu) = -\frac{T}{V} \, \mathbf{Tr}\, \log S^{-1} \,+\, \frac{1}{V}\int\limits_V d^3x \, \frac{(M(\mathbf{x})-m)^2}{4G} \,,
\label{eq:Omega}
\eeq
where 
\beq
S^{-1}(x) = i\gamma^\mu \partial_\mu + \mu\gamma^0 - M(\mathbf{x})
\eeq
is the inverse dressed quark propagator, 
$V$ is a quantization volume, 
and the functional trace $\mathbf{Tr}$ runs over the Euclidean 4-volume $V_4 = [0,\frac{1}{T}] \times V$ 
as well as over Dirac, color and flavor degrees of freedom.

 \subsection{Ginzburg-Landau analysis of critical and Lifshitz points}
 
In order to find the thermodynamically favored state at given $T$ and $\mu$, 
$\Omega$  has to be minimized with respect to the mass function $M(\mathbf{x})$.
This is obviously a nontrivial task. 
The position of the (P)LP, on the other hand, can be determined within a GL analysis, which is possible without knowing the explicit form of 
$M(\mathbf{x})$.

In the chiral limit, $m=0$, the GL expansion corresponds to an expansion of the thermodynamic potential about the chirally restored
phase, $M(\mathbf{x})=0$ in terms of powers and gradients of $M$:
\beq
       \Omega[M] = \Omega[0] + \frac{1}{V} \int d^3x \Big( \alpha_2\,M^2(\x) + \alpha_{4,a}\,M^4(\x)   + \alpha_{4,b}\,(\nabla M(\x))^2
       + \dots     
       \Big)\,.
\label{eq:Omega_GL_cl}
\eeq
Here odd powers are forbidden by chiral symmetry, which implies that $\Omega$ is invariant under $M \rightarrow -M$.
The GL coefficients $\alpha_i$ are functions of $T$ and $\mu$ and thus determine the phase structure.

\begin{figure}[t]
\centering
\includegraphics[width=.19\textwidth]{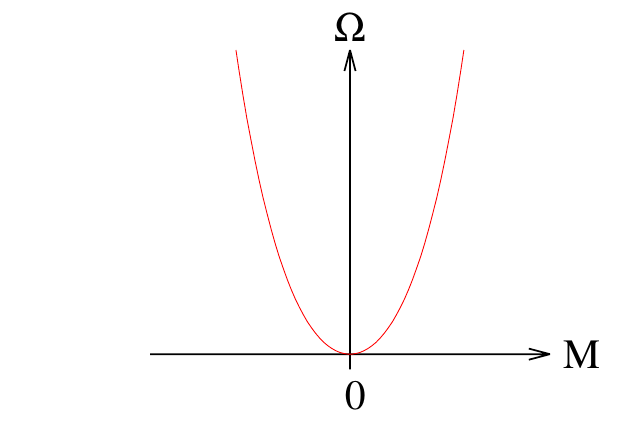}
\includegraphics[width=.19\textwidth]{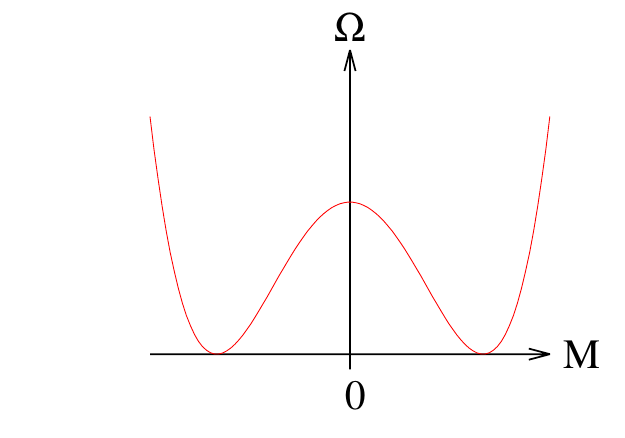}
\includegraphics[width=.19\textwidth]{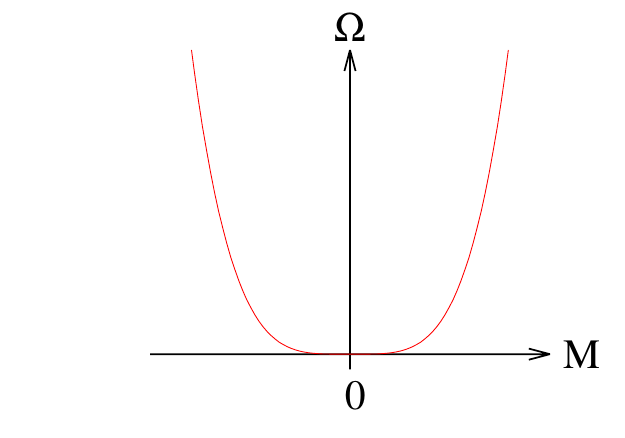}
\includegraphics[width=.19\textwidth]{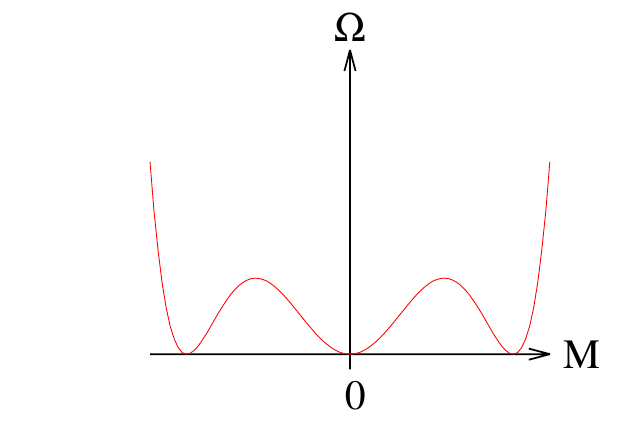}
\includegraphics[width=.19\textwidth]{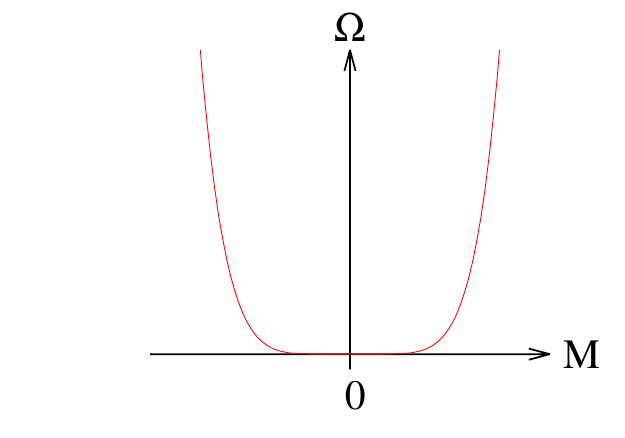}
\caption{ Qualitative behavior of the thermodynamic potential $\Omega$ in the chiral limit as a function of a spatially 
homogeneous order parameter $M$ (from left to right):
(i) restored phase, (ii) broken phase, (iii) second-order phase transition, (iv) first-order phase transition, (v) TCP.
}
\label{fig:Omegaschem0}
\end{figure}

In the following we assume that $M$ and $\nabla M$ are small and that the coefficients of all higher-order terms (indicated
by the ellipsis in \Eq{eq:Omega_GL_cl}) are positive. 
If $\alpha_{4,b}$ is positive as well, gradients are suppressed and the situation reduces to the
standard case of a GL analysis for homogeneous phases, which is illustrated in Fig.~\ref{fig:Omegaschem0}: 
If all coefficients are positive, the minimum of $\Omega$ is the restored phase, $M=0$ (i), while for $\alpha_2<0$ a solution
with $M \neq 0$ ist favored (ii). If $\alpha_{4,a} > 0$ we thus have a second-order phase transition at $\alpha_2=0$ (iii).
For  $\alpha_{4,a} < 0$, on the other hand, there can be a minimum with $M \neq 0$, even for $\alpha_2>0$, so that we can
have a first-order phase transition in this case (iv).  The TCP, where the first-order phase boundary goes over into a 
second-order one is thus given by the condition $\alpha_2 = \alpha_{4,a} = 0$ (v). 

For $\alpha_{4,b} < 0$ inhomogeneous phases can become favored over homogeneous ones.
Its boundary to the restored phase is found to be second order and determined by a balance between the 
free-energy gain due to the negative $\alpha_{4,b}$ term and a free-energy loss caused by a positive $\alpha_2$.
As a consequence, while the  amplitude of the space-dependent modulation vanishes on the phase boundary, 
its wave number stays finite and goes to zero only at the LP 
where both $\alpha_2$ and $\alpha_{4,b}$ vanish.

Away from the chiral limit, i.e., for $m \neq 0$ the situation is more complicated since there is no exactly restored phase. 
We therefore expand the thermodynamic potential about an a priori unknown mass $M_0$, which may depend on $T$ and
$\mu$ but which we assume to be constant in space. 
Then, writing $M(\mathbf{x}) = M_0 + \delta M(\mathbf{x})$
and assuming that the fluctuations $\delta M(\mathbf{x})$ and 
their gradients are small, the expansion reads
\beq
       \Omega[M] = \Omega[M_0] + \frac{1}{V} \int d^3x \left(  \alpha_1 \delta M
        + \alpha_2 \delta M^2 + \alpha_3 \delta M^3
       + \alpha_{4,a} \delta M^4  + \alpha_{4,b} (\nabla \delta M)^2
       + \dots     
       \right) \,,
\label{eq:Omega_GL}       
\eeq
with coefficients $\alpha_i$ which now depend on $T$, $\mu$, and $M_0$.
In contrast to the chiral limit,  the integrand now contains both even and odd powers of $\delta M$.
In the following we will assume that $M_0$ corresponds to a stationary point of the thermodynamic potential
at given values of $T$ and $\mu$. 
This implies that the linear term vanishes, $\alpha_1(T,\mu;M_0) = 0$, which corresponds to a gap equation
for $M_0(T,\mu)$. As we will see below, it is however crucial to keep the $\alpha_3$ term.

\begin{figure}[t]
\centering
\includegraphics[width=.24\textwidth]{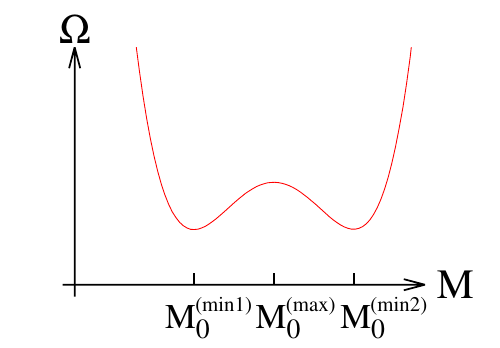}
\includegraphics[width=.24\textwidth]{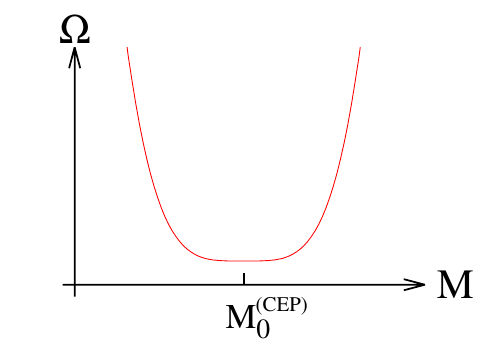}
\includegraphics[width=.24\textwidth]{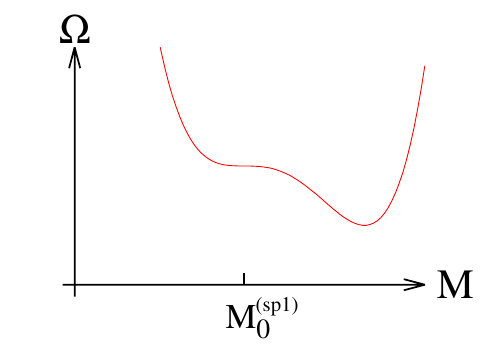}
\includegraphics[width=.24\textwidth]{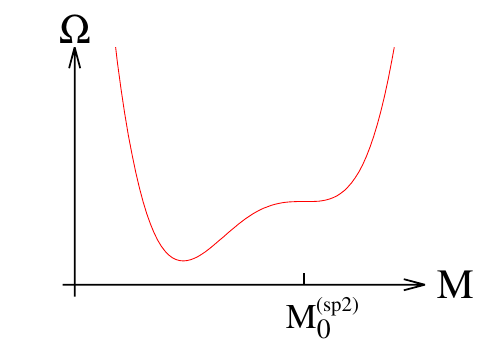}
\caption{Qualitative behavior of the thermodynamic potential $\Omega$ for $m\neq 0$ as a function of a spatially 
homogeneous order parameter $M$ (from left to right):
(i) at a first-order phase boundary , (ii) at the CEP,  (iii) at the left spinodal, (iv) at the right spinodal. }
\label{fig:Omegaschem}
\end{figure}

Again, we first consider the homogeneous case, illustrated in Fig.~\ref{fig:Omegaschem}.
Although there is no strictly restored phase for $m\neq 0$ we can have a first-order phase transition where $\Omega$
as a function of $M$ has two degenerate minima separated by a maximum (i). 
If we now move along the phase boundary towards the CEP, the maximum gets more and more shallow until the three extrema
merge to a single minimum at the CEP (ii). Alternatively we can approach the CEP along the spinodals, which correspond to the
lines in the $T$-$\mu$ plane where one of the two minima merges with the maximum to a saddle point, see  
Fig.~\ref{fig:Omegaschem} (iii) and (iv). Either way one can convince oneself that the CEP is characterized by 
$\alpha_2 = \alpha_3 = 0$. 
For instance on the saddlepoint of the left spinodal (iii) the second derivative of $\Omega$ with respect to $\delta M$ vanishes
while the third derivative is negative, corresponding to $\alpha_2 = 0$ and $\alpha_3 < 0$ if we expand about this point.
On the other hand, if we expand about the saddlepoint of the right spinodal, we get $\alpha_2 = 0$ and $\alpha_3 > 0$,
and therefore both $\alpha_2$ and $\alpha_3$ vanish at the CEP where the two spinodals meet. 

Lacking a chirally restored phase away from the chiral limit, there also cannot exist a Lifshitz point, where the 
second-order boundary between the homogeneous chirally broken and restored phases meets the boundaries of an
inhomogeneous phase. It is still possible however to realize a second-order transition between a homogeneous
and an inhomogeneous phase where the amplitude of the {\it oscillating} part of $M(\x)$ goes to zero, while its
wave number can stay finite. In analogy to the LP we then define the PLP as the point on that phase boundary 
where this wave number also vanishes.
Since our expansion point $M_0$ is constant, the oscillating part is entirely contained in $\delta M$.
In the same way as for the LP in the chiral limit we therefore find that $\alpha_2 = \alpha_{4,b} = 0$ at the PLP.
Moreover, as argued in Ref.~\cite{Buballa:2018hux}, the phase boundary cannot continue in a smooth way beyond this point.
It is therefore most plausible to identify it with the tip of the inhomogeneous phase in the $T$-$\mu$ plane.
This was also confirmed numerically. 

In summary, we find
\beq
      \alpha_2 = \alpha_{4,a} = 0  \quad \text{at the TCP,} \qquad 
      \alpha_2 = \alpha_{4,b} = 0 \quad \text{at the LP}
\label{eq:TCPLP}
\eeq      
in the chiral limit for an expansion around $M=0$,  
and 
\beq
       \alpha_2 = \alpha_{3} = 0  \quad \text{at the CEP,} \qquad
      \alpha_2 = \alpha_{4,b} = 0 \quad \text{at the PLP}
\eeq
away from the chiral limit 
for an expansion around a stationary point $M_0(T,\mu)$, which is obtained by simultaneously solving 
the gap equation $\alpha_1 = 0$.

\subsection{Determination of the GL coefficients}
\label{sec:GLcoeffs2}

For the explicit determination of the GL coefficients in the NJL model we basically follow Ref.~\cite{Nickel:2009ke}.
Inserting our decomposition $M(\x) = M_0 + \delta M(\x)$ of the constituent mass function into 
the mean-field thermodynamic potential, \Eq{eq:Omega}, one gets
\beq
\Omega = -\frac{T}{V} \, \mathbf{Tr}\, \log(S_0^{-1}-\delta M) 
                \,+\, \frac{1}{V}\int\limits_V d^3x \, \frac{(M_0-m+\delta M(\x))^2}{4G} \,,
\eeq
where $S_0^{-1}(x) = i\gamma^\mu \partial_\mu + \mu\gamma^0 - M_0$ depends only on the constant mass $M_0$.
Expanding the logarithm into a Taylor series about $S_0^{-1}$ this can be written as 
$\Omega = \sum_{n=0}^\infty \Omega^{(n)}$
where $\Omega^{(n)}$ is of the $n$th order in the fluctuating fields $\delta M$.
Specifically one obtains 
\begin{alignat}{1}
       \Omega^{(1)} &=  \frac{T}{V} \, \mathbf{Tr}\,\left( S_0 \delta M \right)  
       \,+\, \frac{M_0-m}{2G}\, \frac{1}{V}\int\limits_V d^3x \; \delta M(\x) \,,
\\
       \Omega^{(2)} &= \frac{1}{2} \frac{T}{V} \, \mathbf{Tr}\,\left( S_0 \delta M \right)^2 
       \,+\, \frac{1}{4G}\, \frac{1}{V}\int\limits_V d^3x \; \delta M^2(\x) \,,
\\
       \Omega^{(n)} &= \frac{1}{n} \frac{T}{V} \, \mathbf{Tr}\,\left( S_0 \delta M \right)^n        \quad \text{for } n\geq 3.
\end{alignat}
The functional traces are given by
\beq
       \mathbf{Tr}\,\left( S_0 \delta M \right)^n = 2 N_c \int \prod_{i=1}^n d^4x_i  \, \mathrm{tr}_\mathrm{D} \left[
       S_0(x_n,x_1) \delta M(\x_1)S_0(x_1,x_2) \delta M(\x_2) \dots S_0(x_{n-1},x_n) \delta M(\x_n)\right]\,,
\eeq
where the integrals are again over $V_4$,
$\mathrm{tr}_\mathrm{D}$ indicates a trace in Dirac space, and
we have already turned out the trivial traces in color and flavor space.
Noting that $S_0$ is the standard propagator of a free fermion with mass $M_0$ at chemical potential $\mu$,
the evaluation of the Dirac trace is straightforward using the momentum-space representation of $S_0$.
After performing a gradient expansion of $\delta M(\x_i)$ about $\x_1$, i.e.,
$\delta M(\x_i) = \delta M(\x_1) + \nabla \delta M(\x_1)\cdot(\x_i - \x_1) + \dots$,  
one can also perform the integrations over all space-time variables $x_i \neq x_1$ 
and then compare the results with \Eq{eq:Omega_GL} to read off the GL coefficients. 
One finds: 
\begin{alignat}{1}
\alpha_1 & = \frac{M_0-m}{2G} + M_0 F_1 \,,
\\
\alpha_2 & = \frac{1}{4G} + \frac{1}{2} F_1 +  M_0^2 F_2 \,,
\\
\alpha_3 & = M_0 \left( F_2 +  \frac{4}{3} M_0^2 F_3 \right) \,,
\\
\alpha_{4,a} & = \frac{1}{4} F_2 +  2  M_0^2 F_3 + 2 M_0^4 F_4 \,,
\\
\alpha_{4,b} & = \frac{1}{4} F_2 +  \frac{1}{3}  M_0^2 F_3 \,,
\end{alignat}
where we have introduced the functions
\beq
       F_n = 8 N_c \int \frac{d^3p}{(2\pi)^3}\, T\sum\limits_j \frac{1}{[(i\omega_j +\mu)^2 - \p^2 -M_0^2]^{n}} 
\label{eq:Fn}
\eeq
with fermionic Matsubara frequencies $\omega_j = (2j+1)\pi T$.

Even without further evaluation, we can spot several interesting consequences of the above results:
\begin{itemize}
\item In the chiral limit, $m=0$, the restored phase $M_0 = 0$ is a solution of the gap equation $\alpha_1 = 0$.
         Expanding about this solution, $\alpha_3$ vanishes as well, in agreement with \Eq{eq:Omega_GL_cl}. 
         Moreover, we reproduce the result $\alpha_{4,a} = \alpha_{4,b}$ of Ref.~\cite{Nickel:2009ke}, 
         meaning that the LP coincides with the TCP in the chiral limit. 

\item Considering $M_0 \neq 0$, but taking the limit $M_0 \rightarrow 0$ the $\alpha_3 = 0$ line in the $T$-$\mu$ plane
         approaches the $\alpha_{4,a} = 0$ line, and hence the CEP converges against the TCP (coinciding with the LP).
         
\item  For arbitrary $M_0$ we have $\alpha_3 = 4M_0 \alpha_{4,b}$. For $M_0\neq 0$ this implies that the PLP
          coincides with the CEP. 
\end{itemize}
Hence, since the PLP corresponds to the tip of the inhomogeneous phase, the latter ``ends'' exactly at the same point 
as the first-order phase boundary in a purely homogeneous treatment of the same model. This is analogous to the 
coincidence of the LP with the TCP in the chiral limit.
Of course, the GL study does not prove the existence of the CEP in the first place but only predicts that {\it if} there is
a first-order phase transition in a homogeneous treatment of the model, then there should be an inhomogeneous phase, 
at least in the vicinity of the CEP. 
In fact, numerically one finds that the inhomogeneous phase covers the entire first-order boundary.
We also found that the inhomogeneous phase shrinks as one moves away from the chiral limit, but survives even 
at significantly large values of the current quark mass \cite{Buballa:2018hux}.

On the other hand, even in the chiral limit the LP and TCP do not necessarily coincide anymore, if the model is modified,
e.g., by adding a vector term \cite{Carignano:2010ac}. 
In the next section we 
will study whether this is the case when strange quarks are added to the model.

\section{Including strange quarks}
\label{sec:strange}

We now consider the Lagrangian
\beq
\mathcal{L} = \bar\psi\left( i\gamma^\mu \partial_\mu - \hat m \right) \psi 
+ \mathcal{L}_4 + \mathcal{L}_6
\eeq
where $\psi = (u,d,s)^T$ denotes a quark field with three flavor degrees of freedom
and $\hat m$ is the corresponding bare mass matrix. 
The last two terms describe a $U(3)_L\times U(3)_R$ invariant four-point interaction
\beq
\mathcal{L}_4 
=  G \sum_{a=0}^8\left[ (\bar\psi\tau_a\psi)^2 + (\bar\psi i\gamma_5 \tau_a \psi)^2 \right],
\eeq
and a six-point (``Kobayashi - Maskawa - 't Hooft'', KMT) interaction 
\beq
\mathcal{L}_6 = -K \left[ \mathrm{det}_f \bar\psi(1+\gamma_5)\psi +  \mathrm{det}_f  \bar\psi(1-\gamma_5)\psi \right] \,,
\eeq
which is $SU(3)_L\times SU(3)_R$ symmetric but breaks the $U(1)_A$ symmetry,
mimicking the axial anomaly.
In the former $\tau_a$, $a=1,\dots,8$, denote Gell-Mann matrices in flavor space while 
$\tau_0 = \sqrt{2/3} \,\mathbb{1}$ is proportional to the unit matrix.

Starting from these Lagrangians, we perform again a mean-field approximation, considering the 
non-strange and  strange condensates
\beq
       \sigma_\ell(\x)  \equiv \ave{\bar u u}(\x) =  \ave{\bar d d}(\x)       
       \quad \text{and} \quad
       \sigma_s(\x)  \equiv \ave{\bar s s}(\x)\,,       
\eeq 
which may be space dependent.
Proceeding analogously as in the two-flavor case, the mean-field thermodynamic potential becomes
\beq
\Omega(T,\mu) = -\frac{T}{V} \, \mathbf{Tr}\, \log S^{-1} \,
+\, \frac{1}{V}\int\limits_V d^3x \, \big\{ 2G (2\sigma_\ell^2 + \sigma_s^2) -4K \sigma_\ell^2 \sigma_s \big\} \,,
\label{eq:Omega3}
\eeq
where the inverse dressed quark propagator is now given by
$S^{-1} = \mathrm{diag}_f(S^{-1}_u,S^{-1}_d,S^{-1}_s)$
with the flavor components
\beq
       S^{-1}_f(x) =  i\gamma^\mu \partial_\mu + \mu\gamma^0 - M_f(\mathbf{x}) \,.
\eeq
and the constituent mass functions
\beq
       M_\ell(\x) = m_\ell - 4G \sigma_\ell(\x) + 2K\sigma_s(\x)  \sigma_\ell(\x) 
       \quad \text{and}\quad
       M_s(\x) = m_s - 4G \sigma_s(\x) + 2K \sigma_\ell^2(\x)\,.
\eeq
Here we assumed isospin symmetry in the light sector, $m_u = m_d \equiv m_\ell$, and thus $M_u = M_d \equiv M_\ell$.

From the above expressions we can see that the different flavors only couple through the six-point interaction 
and decouple for $K=0$. This is well known from earlier studies of this model 
but will also play an important role in our analysis below.

\subsection{Ginzburg-Landau analysis}
 
The GL expansion of the thermodynamic potential introduced above is again complicated by the fact that in the presence 
of nonvanishing bare quark masses there is no chirally restored solution. 
On the other hand, at least for the strange quark, neglecting the mass would be a rather unrealistic approximation.
We therefore consider a partially simplified problem with $m_s\neq 0$ but $m_\ell=0$. 
In this case a two-flavor restored solution $\sigma_\ell = 0$ exists, which we take as the expansion point of our GL analysis.
In the strange-quark sector we proceed similarly as in the previous section and expand about a 
$T$ and $\mu$ dependent but spatially constant  condensate $\sigma_s^{(0)}$, corresponding to a stationary point of $\Omega$
at $\sigma_\ell = 0$. We thus write
\beq
       \sigma_s(\x) = \sigma_s^{(0)} + \delta\sigma_s(\x)
\eeq
and expand
\beq
       \Omega[\sigma_\ell,\sigma_s]
        =  \Omega[0,\sigma_s^{(0)}] +  \frac{1}{V} \int d^3x \, \omega_{GL}(\Delta_\ell,\Delta_s) \,,
\eeq
where
\beq
       \Delta_\ell(\x) = -4G \sigma_\ell(\x) \quad \text{and}\quad \Delta_s(\x) = -4G \delta\sigma_s(\x) 
\eeq
are proportional to the fluctuations. 
The GL function takes the form
\begin{alignat}{1}
\omega_{GL}(\Delta_\ell,\Delta_s)
= \phantom{+} &\alpha_2 \Delta_\ell^2 + \alpha_{4,a} \Delta_\ell^4 + \alpha_{4,b} (\nabla\Delta_\ell)^2 + \dots
\nonumber\\
+ &\beta_1 \Delta_s  + \beta_2 \Delta_s^2  + \beta_3 \Delta_s^3 + \beta_{4,a} \Delta_s^4 + \beta_{4,b} (\nabla\Delta_s)^2
+ \dots
\nonumber\\
+ &\gamma_3 \Delta_\ell^2 \Delta_s + \gamma_4 \Delta_\ell^2 \Delta_s^2  + \dots \,,
\label{eq:Omega_GL_s}
\end{alignat}
where the $\alpha$- and $\beta$-terms correspond to the contributions from the light and strange condensates, respectively, 
and the $\gamma$ terms describe the mixing.
Since we expand about the two-flavor restored phase, only even powers of $\Delta_\ell$ are allowed, 
and the structure of the $\alpha$-terms is the same as in the GL expansion of the two-flavor model in the chiral limit, 
\Eq{eq:Omega_GL_cl}.
For $\Delta_s$, on the other hand, we can also have odd terms, so that the $\beta$-terms have a structure as in 
\Eq{eq:Omega_GL}. 
Again, for a stationary expansion point, the linear coefficient $\beta_1$ has to vanish, giving rise to a gap equation for
$\sigma_s^{(0)}$ at given $T$ and $\mu$. 

For vanishing $\gamma_i$  the GL analysis of the non-strange sector would be analogous to the 
two-flavor case in the chiral limit, i.e., the tricritical and Lifshitz points are related to the $\alpha$ coefficients as in 
\Eq{eq:TCPLP}. In order to study how these relations get modified by $\gamma_i \neq 0$, 
we eliminate $\Delta_s$ by extremizing the thermodynamic potential with respect to this function.
To this end we employ the Euler-Lagrange equations,
\beq
      \frac{\partial \omega_{GL}}{\partial \Delta_s} -\partial_i \frac{\partial \omega_{GL}}{\partial \partial_i\Delta_s} = 0\,,
\eeq
which yields
\beq
       \Delta_s = -\frac{\gamma_3}{2\beta_2} \Delta_\ell^2 + \dots  \equiv   \Delta_s^\mathit{extr}\,.
\label{eq:Dselim}       
\eeq
Here we have already used the gap equation $\beta_1=0$. 
The ellipsis indicates higher orders in $\Delta_\ell$ and gradients, which we treat equally,  
$\mathcal{O}(\nabla^n)=\mathcal{O}(\Delta_\ell^n)$. 
From the above equation we can then see that $\Delta_s$ is of the order $\mathcal{O}(\Delta_\ell^2)$.
Inserting \Eq{eq:Dselim} into \Eq{eq:Omega_GL_s} and keeping only terms up to the order $\mathcal{O}(\Delta_\ell^4)$,
one then obtains
\beq
       \omega_{GL}(\Delta_\ell,\Delta_s^\mathit{extr}) 
       = \alpha_2 \Delta_\ell^2 + \left(\alpha_{4,a} - \frac{\gamma_3^2}{4\beta_2} \right) \Delta_\ell^4 
       + \alpha_{4,b} (\nabla\Delta_\ell)^2 + \dots
\eeq
We thus find that the quartic term in $\Delta_\ell$ gets an additional contribution through the coupling to the strange quarks,
while the gradient term does not. 
So, instead of \Eq{eq:TCPLP}, we now have
\beq
      \alpha_2 = \alpha_{4,a}  - \frac{\gamma_3^2}{4\beta_2} = 0  \quad \text{at the TCP,} \qquad 
      \alpha_2 = \alpha_{4,b} = 0 \quad \text{at the LP}\,,
\label{eq:TCPLP3}
\eeq      
and therefore, even if $\alpha_{4,a}$ and $\alpha_{4,b}$ were still equal (as they are in the two-flavor model),
the TCP and the LP would no longer coincide for $\gamma_3\neq 0$. 

Of course, in order to be able to make definite statements we have to evaluate the relevant GL coefficients.
This can be done in the same way as discussed in Sec.~\ref{sec:GLcoeffs2} for the two-flavor model.
We find
\begin{alignat}{1}
\alpha_2 & = (1+2\delta)\frac{1}{4G} +(1+\delta)^2 \frac{1}{2} F_1^{(\ell)} +  \frac{K}{16G^2}M_{s,0} F_1^{(s)}\,,
\\
\alpha_{4,a} & = (1+\delta)^4 \frac{1}{4} F_2^{(\ell)} +  \frac{K^2}{256G^4}\left(F_1^{(s)}   +2M_{s,0}^2 F_2^{(s)} \right)\,,
\\
\alpha_{4,b} & =  (1+\delta)^2 \frac{1}{4} F_2^{(\ell)} \,,
\\
\beta_2 & = \frac{1}{8G} + \frac{1}{4} F_1^{(s)} +  \frac{1}{2} M_{s,0}^2 F_2^{(s)}\,,
\\
\gamma_3 & = \frac{K}{2G^2} \left\{ \frac{1}{8G} + (1+\delta) \frac{1}{4} F_1^{(\ell)} + \frac{1}{8} F_1^{(s)} 
                                                        +  \frac{1}{4} M_{s,0}^2 F_2^{(s)}\right\}\,,
\end{alignat}
where $M_{s,0} = m_s - 4G \sigma_s^{(0)}$, and
$F_n^{(\ell)}$ and $F_n^{(s)}$ are the functions defined in \Eq{eq:Fn} with $M_0 = 0$ and $M_0 = M_{s,0}$, respectively. 
Furthermore we defined $\delta = -\frac{K}{2G}\sigma_s^{(0)}$.

As we have seen earlier, the different flavors are only coupled through the six-point interaction. 
Indeed, for $K=0$ and thus $\delta=0$, the flavor mixing coefficient $\gamma_3$ vanishes,
and the $\alpha$ coefficients  reduce to the corresponding two-flavor expressions in the chiral limit. 
In particular, we reproduce again the result of Ref.~\cite{Nickel:2009ke} that TCP and LP coincide in this case. 

For $K\neq0$, on the other hand, $\gamma_3\neq0$ and therefore TCP and LP would not even coincide 
if $\alpha_{4,a}$ and  $\alpha_{4,b}$ were equal,  as seen in \Eq{eq:TCPLP3}. 
Moreover, $\alpha_{4,a}$ and  $\alpha_{4,b}$ are {\it not} equal, and the two effects are not found to cancel each other. 
We thus find that TCP and LP split for $K\neq 0$, i.e., as a consequence of the axial anomaly. 
A quantitative investigation of this effect will be presented in a forthcoming paper \cite{inprep}.

\section{Conclusions}

In this contribution we discussed how the introduction of nonzero bare quark masses and of strange quarks affects
inhomogeneous phases in the NJL model. 
To this end we performed a Ginzburg-Landau analysis, which allows to investigate these effects
without specifying the shape of the spatial modulations and is valid close to the LP.
Strictly speaking, away from the chiral limit one cannot define a LP as the location where the 
second-order boundary between the homogeneous chirally broken and restored phases meets the boundaries of an
inhomogeneous phase. We therefore
introduce the pseudo-LP as the point on that phase boundary where both the amplitude and the wave number
of the spatially modulated part of the chiral order parameter vanish.
Our results for a two-flavor model show that the PLP exactly coincides with the CEP \cite{Buballa:2018hux}, 
supporting the numerical evidence found in \cite{Nickel:2009wj}.

When extending our analysis to include 
strange quarks we find on the other hand
that 
the LP and the TCP split
when the different flavors are coupled through a non-vanishing KMT vertex. 
Here we have considered massive strange quarks but massless up and down quarks. 
In a forthcoming publication we will perform a full quantitative investigation 
of this effect and corroborate it with numerical results away from the chiral limit. 
There we will also investigate the behavior of the LP  for very small values of $m_s$. 
This is of special interest, as 
we expect that in this limit the TCP eventually reaches the $T$-axis. If the LP followed the same behavior, it would mean that 
it would be possible to realize an inhomogeneous phase all the way to the regime of vanishing densities, which, if also realized in QCD,
could be investigated on the lattice.
In the light of the above results we conclude that this scenario 
 might however not be
realized, as in the three-flavor case the LP does not coincide with the TCP anymore. 
A quantitative investigation of this effect is therefore required. 

Finally, we recall that,
while our analysis focused on the lowest-order GL coefficients which are relevant for pinpointing the location of the CP and the (P)LP, 
it would also be interesting to work out higher-order
coefficients, as they can give informations on the favored shape of the spatially modulated chiral condensate \cite{Abuki:2011pf,Carignano:2018hvn}, as well as determine the phase boundary to the chirally restored phase \cite{Carignano:2017meb}.

\subsection*{Acknowledgments}
We acknowledge support by the Deutsche Forschungsgemeinschaft (DFG, German Research Foundation) through the CRC-TR 211 `Strong-interaction matter under extreme conditions' - project number 315477589 - TRR 211.
S.C.\ has been supported by the projects FPA2016-81114-P and FPA2016-76005-C2-1-P (Spain), and by the project 2017-SGR-929 (Catalonia).

\input{QCHS.bbl}

%\bibliographystyle{JHEP}
%\bibliography{biblio1118.bib}

%\begin{thebibliography}{99}
%\bibitem{...}
%....
%
%\end{thebibliography}

\end{document}

%% file: QCHS.bbl
\providecommand{\href}[2]{#2}\begingroup\raggedright\endgroup